\documentclass{article}
\usepackage{epsfig}
\date{}
\title{New heavy charged leptons at future 
high energy electron-positron colliders.}
\author{F.M.L. Almeida Jr., Y. A. Coutinho, \\
J. A. Martins Sim\~oes, S. Wulck,\\
Instituto de F\'\i isica, Universidade Federal do Rio de
Janeiro, \\
Rio de Janeiro, RJ, Brazil, \\
M. A. B. do Vale,\\
Departamento de Ci\^encias Naturais, \\
Universidade Federal de S\~ao Jo\~ao del Rei, Brazil}

\begin{document}
\maketitle

\begin{abstract}
New heavy charged lepton production and decay signatures at future electron-positron colliders are investigated at $\sqrt {s}=500$ GeV. The consequences of model dependence for vector singlets and vector doublets are studied. Distributions are calculated including hadronization effects and experimental cuts that suppress the standard model background. The final state leptonic energy distributions are shown to give a very clear signature for heavy charged leptons.
\end{abstract}

\section{Introduction}
The possibility of new heavy charged leptons is present in many extentions of the Standard 
Model (SM). In grand unified theories beyond SU(5) we have new fermionic fields \cite{GOD}. In superstring theories with an $E_6$ gauge group after compactification, the fermions in the fundamental 27 representation have new leptonic doublets \cite{HEW} and new neutral lepton singlets \cite{GON}. In left-right symmetric models \cite{MUR} we have also new mirror fermions that are expected to be much heavier than the presently known fermions from the SM. Extended electroweak gauge groups with three anomaly free families \cite{VIC} contain new fermions. We can also have a new  fourth family, similar to the electron-muon-tau families, but with much higher mass states.
\par
It is well known that experimentally no clear signal of such fermions was found so far \cite{PDG}. If these particles exist, the latest experimental limits imply that they must have a mass greater than $100$ GeV. However, the possibility of generalizing the SM into a deeper model is a strong motivation for both experimental searches at the next colliders as well as for theoretical studies. The recent experimental confirmation of light neutrino masses and oscillations \cite{ZUB} also raises the question of new possible heavy neutral leptons. These neutral heavy leptons could be responsible for the see-saw mechanism for small neutrino masses. As the presently known leptons appear in neutral and charged states, it is reasonable to suppose that new heavy states will show the same behavior. The new heavy charged leptons could be searched for in future high energy hadron-hadron colliders \cite{HAD}, lepton-hadron colliders \cite{LHA}, lepton-lepton colliders \cite{LEL} and photon-electron \cite{PHO} colliders. In this paper we will concentrate our attention on the electron-positron option at $ \sqrt {s}= 500$ GeV. This will be the first energy option for 
the linear collider at NLC \cite{NLC} as well as for the TESLA \cite{TES} project at DESY before they move to higher center of mass energies and luminosities. Most of our results can be scaled for higher energy options. We have considered an yearly integrated luminosity of $ {\cal {L}}=10^5$pb$^{-1}$.
\par
The new heavy charged and neutral leptons have been studied intensively by many authors   \cite{GOD,HEW,GON,MUR,VIC,PDG,ZUB,HAD,LHA,LEL,PHO,MIX,VIO,PIL} For heavy lepton masses up to $M_{L}=\sqrt {s}/2$ the total cross section is dominated by pair production due to photon exchange. So this mechanism is quite model independent. We can extend the heavy lepton mass range to $\sqrt{s}>M_{L}> \sqrt{s}/2$ by looking for   single heavy lepton production. In this case, and also in the decay of heavy leptons, there is a strong model dependence on the mixing between light and heavy leptons as well as in the heavy-to-heavy lepton couplings.
\par
In the present work we study the dynamical and kinematical properties for production and also for the decay of new heavy charged leptons into presently known particles. These distributions are compared with the expected background coming from the SM and we show how the signal can be clearly separated from the background with a great significance, independent of hadronization effects. Another point discussed  is the model dependence for new heavy charged lepton experimental search. 
\par
This paper is organized as follows. In section 2 we discuss the models adopted and their parameters. In section 3 we present the results from the various processes involving heavy leptons, studied in this paper, comparing the distributions from SM, vector singlet model (VSM) and vector doublet model (VDM). Finally, conclusions are shown in section 4.
\par

\section{The models}
 The high precision data on the SM parameters imply strong limits on new heavy gauge bosons. So, our first hypothesis is that in the next $e^{-}e^{+}$ collider at $ \sqrt{s}= 500$ GeV the mechanism for production and decay of new heavy charged leptons will be dominated by the presently known electroweak gauge bosons. The second hypothesis is that the new heavy leptons can mix with the light leptons in each family. The mixing angle parameters are constrained by the $Z$ properties and by asymmetries in lepton-lepton scattering. They imply a small mixing \cite {MIX}, of the order of $\sin^2 \theta_{mix} < 0.005$ with  $95 \%$ c.l. Throughout this paper we will consider all mixing angles ($\theta^e_{L,R}$ and $\theta^{\nu}_{L,R}$)
between the new and the ordinary leptons (charged and neutral) equal to the present upper bound $\sin^2 \theta_{mix} = 0.005$.
 The next theoretical point concerns the quantum numbers that can be assigned to new heavy leptons. Here we have a number of possibilities, but we will restrict our analysis to only two models: new vector singlets and new vector doublets. For the VSM the new charged heavy lepton (written as $L$) is a singlet, as well as the new neutral lepton (written as N). For the VDM both new heavy leptons are placed in doublets. There are other possible models for new heavy leptons such as mirror and fourth generation models. We have restricted our analysis to the VDM and VSM since they give the more important features of model dependence with fewer parameters than other models.
\par
The interactions are then given by the Lagrangians for neutral and charged currents:

\bigskip
\begin{equation}
{\cal L}_{cn}=-\frac{g}{4c_W}Z_{\mu}\overline{\psi_i}\gamma^{\mu}\left(g_V^{ij}-g_A^{ij}\gamma_5\right)\psi_j + h.c.
\end{equation}
\bigskip
\begin{equation}
{\cal L}_{cc}=-\frac{g}{2\sqrt{2}}W_{\mu}\overline{\psi_i}\gamma^{\mu}\left(g_V^{ij}-g_A^{ij}\gamma_5\right)\psi_j + h.c.
\end{equation}
where {\it i} and  {\it j} are the possible combinations of light and heavy leptons. The couplings depend on mixing parameters that are given in table 1 for the VSM, and in table 2 for the VDM. In the next section we will discuss the role of heavy neutrinos in the heavy charged lepton decays. In tables 1 and 2, we present the couplings for the case of new heavy neutrino states of the Dirac type. In this case  a global family symmetry is present and no lepton number violation occurs.\par
\begin{center}
\begin{table}
\caption{Vector singlet model couplings between bosons and fermions with N as a Dirac field}
\begin{tabular}{|c|c|c|}\hline
 Vertice & $g_V$  & $g_A$  \\ \hline
$Z \rightarrow e^+ e^-$ & $-\cos^2 \theta^e_L + 4s_W^2$ & $-\cos^2 \theta^e_L$ \\ \hline
$Z \rightarrow \nu_e \bar\nu_e$ & $\cos^2 \theta^{\nu}_L$ &  $\cos^2 \theta^{\nu}_L$ \\ \hline
$Z \rightarrow e^+ L_e^-$ & $-\cos\theta^e_L \sin\theta^e_L$ & $-\cos\theta^e_L \sin\theta^e_L$ \\ \hline
$Z \rightarrow \nu_e \bar N$ &  $\cos\theta^{\nu}_L \sin\theta^{\nu}_L$  & $\cos\theta^{\nu}_L \sin\theta^{\nu}_L$ \\ \hline
$Z \rightarrow N \bar N$ & $\sin^2 \theta^{\nu}_L$ & $\sin^2 \theta^{\nu}_L$\\ \hline
$Z \rightarrow L_e^+ L_e^- $ & $-\sin^2 \theta^e_L + 4s_W^2$ & $-\sin^2 \theta^e_L$ \\ \hline\hline
% & & \\ \hline
$W^+ \rightarrow \nu_e e^+$ & $\cos\theta^{\nu}_L \cos\theta^e_L$ & $\cos\theta^{\nu}_L \cos\theta^e_L$ \\ \hline
$W^+ \rightarrow \bar N e^+$ & $\sin\theta^{\nu}_L \cos\theta^e_L$ & $\sin\theta^{\nu}_L \cos\theta^e_L$ \\ \hline
$W^+ \rightarrow  \nu_e L_e^+$ & $\cos\theta^{\nu}_L \sin\theta^e_L$ & $\cos\theta^{\nu}_L \sin\theta^e_L$ \\ \hline
$W^+ \rightarrow  N L_e^+$ & $\sin\theta^{\nu}_L \sin\theta^e_L$ & $\sin\theta^{\nu}_L \sin\theta^e_L$ \\ \hline 
\end{tabular}
\end{table}
\end{center}

\begin{center}
\begin{table}
\caption{Vector doublet model couplings between bosons and fermions with N as a Dirac field}
\begin{tabular}{|c|c|c|}\hline
 Vertice & $g_V$  & $g_A$  \\ \hline
$Z \rightarrow e^+ e^-$ & $-1 -\sin^2 \theta^e_R + 4s_W^2$ & $-\cos^2 \theta^e_R$ \\ \hline
$Z \rightarrow \nu_e \bar\nu_e$ & $1+\sin^2 \theta^{\nu}_R$ &  $\cos^2 \theta^{\nu}_R$ \\ \hline
$Z \rightarrow e^+ L_e^-$ & $\cos\theta^e_R \sin\theta^e_R$ & $-\cos\theta^e_R \sin\theta^e_R$ \\ \hline
$Z \rightarrow \nu_e \bar N$ &  $-\cos\theta^{\nu}_R \sin\theta^{\nu}_R$  & $\cos\theta^{\nu}_R \sin\theta^{\nu}_R$ \\ \hline
$Z \rightarrow N \bar N$ & $1+\cos^2 \theta^{\nu}_R$ & $\sin^2 \theta^{\nu}_R$\\ \hline
$Z \rightarrow L_e^+ L_e^- $ & $-1-\cos^2 \theta^e_R + 4s_W^2$ & $-\sin^2 \theta^e_R$ \\ \hline\hline
% & & \\ \hline
$W^+ \rightarrow \nu_e e^+$ & $\cos(\theta^{\nu}_L - \theta^e_L)  $ &  $\cos(\theta^{\nu}_L - \theta^e_L) $ \\
                           & $+ \sin\theta^{\nu}_R \sin\theta^e_R$ &  $ - \sin\theta^{\nu}_R \sin \theta^e_R$ \\ \hline
$W^+ \rightarrow \bar N  e^+$ & $\sin(\theta^{\nu}_L - \theta^e_L)$  &  $\sin(\theta^{\nu}_L - \theta^e_L)$ \\
                             & $- \cos\theta^{\nu}_R \sin\theta^e_R$  &  $ + \cos\theta^{\nu}_R \sin \theta^e_R$ \\ \hline
$W^+ \rightarrow \nu_e L_e^+$ & $\sin(\theta^{\nu}_L - \theta^e_L)  $ & $\sin(\theta^{\nu}_L - \theta^e_L)  $ \\ 
                              & $ - \sin\theta^{\nu}_R \cos\theta^e_R$ & $ + \sin\theta^{\nu}_R \cos \theta^e_R$ \\ \hline
$W^+\rightarrow  N L_e^+$ & $\cos(\theta^{\nu}_L - \theta^e_L)  $  &  $\cos(\theta^{\nu}_L - \theta^e_L)  $\\
                          & $ - \cos\theta^{\nu}_R \cos \theta^e_R$  &  $ + \cos\theta^{\nu}_R \cos \theta^e_R$\\ \hline
\end{tabular}
\end{table}
\end{center}

The heavy neutrinos could also be of the Majorana type. Then there are small changes in their couplings due to the identity $\bar \Psi_{N}\gamma^{\mu}\Psi_{N} \equiv 0$, i.e., pure vector couplings are zero. We  have in each family the same couplings  of the heavy Majorana neutrino with ordinary leptons. In the  Majorana  case, we have the possibility that heavy charged lepton decays could also violate lepton number  conservation \cite{VIO} due to the chain,
\begin{eqnarray}
L_{i}\longrightarrow & & N \: W \cr
& & \vert  \cr
& & {}^{\displaystyle \longrightarrow \ell^\pm \: W^\mp}
\end{eqnarray}
where $\ell,i= e,\mu,\tau$,
so that the cross sections are the same for dilepton final states of the type $ e^- e^+ \longrightarrow e^{\pm} + \ell^{\mp} + W^{\pm} + W^{\mp}$. 
\par
In all calculations presented in this paper we have considered $N$ as a heavy Majorana neutrino that couples to all families, with mass $M_N= 100$ GeV. We are assuming the heavy neutral lepton mass to be smaller than the heavy charged lepton mass as in the light lepton mass spectrum for the three known lepton families.
\par
For simplicity we have concentrated on the first (electron) family and our results can be generalized to the other families.

\section {Results}

 In this section we present the main channels, cross sections and distributions, comparing the signal from VSM and VDM  and the SM background. This procedure allow us to obtain a clear signature for heavy charged leptons. As many of these processes involve a large number of diagrams, we have employed the CalcHep \cite{HEP} package. The VSM and VDM were implemented in the unitary and Feynman gauges and both results were carefully checked and compared. So, the results here presented are complete first order calculations.\par

\begin{figure}
\resizebox{1.2\hsize}{!}{\includegraphics*{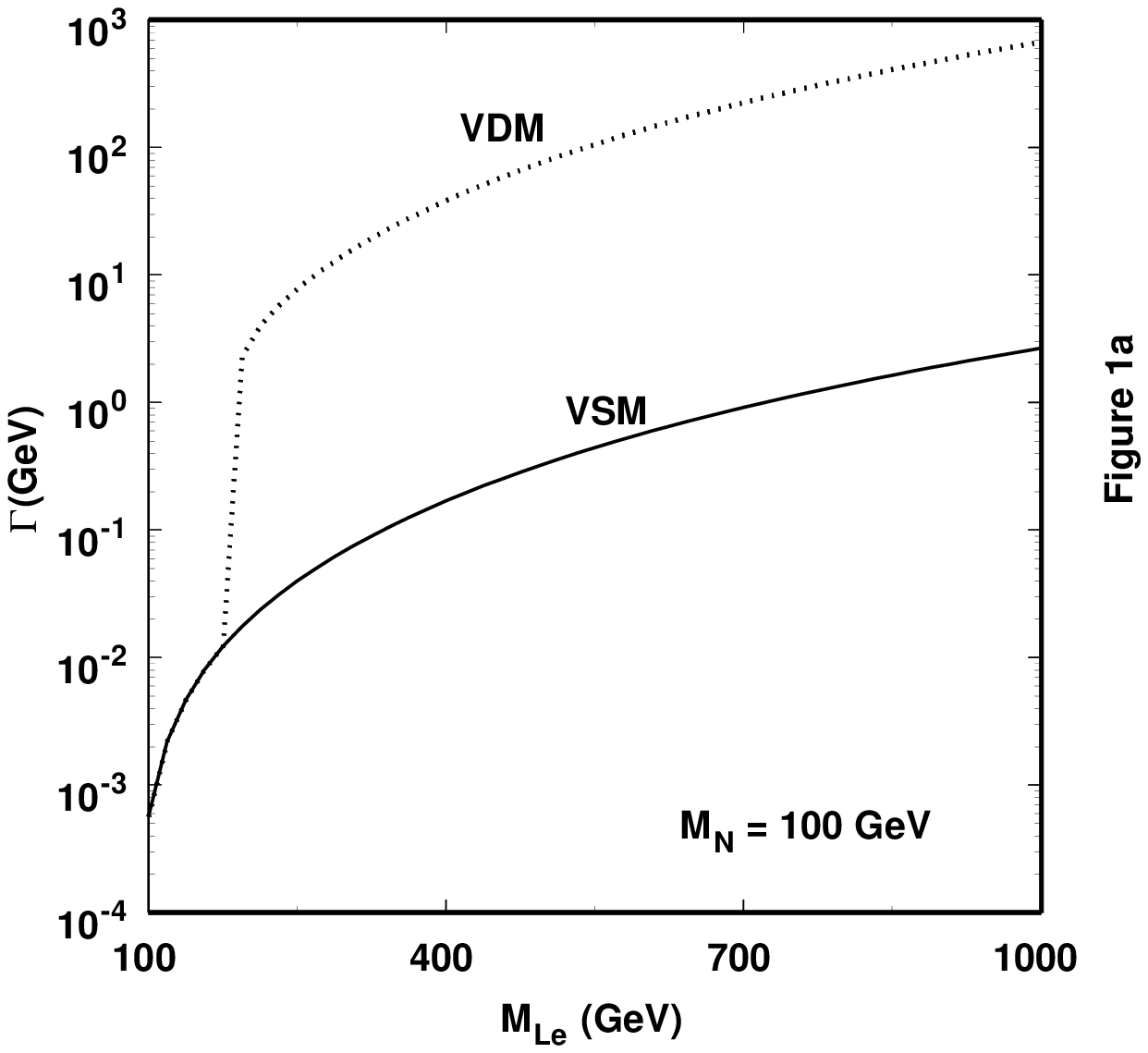}}
\resizebox{1.2\hsize}{!}{\includegraphics*{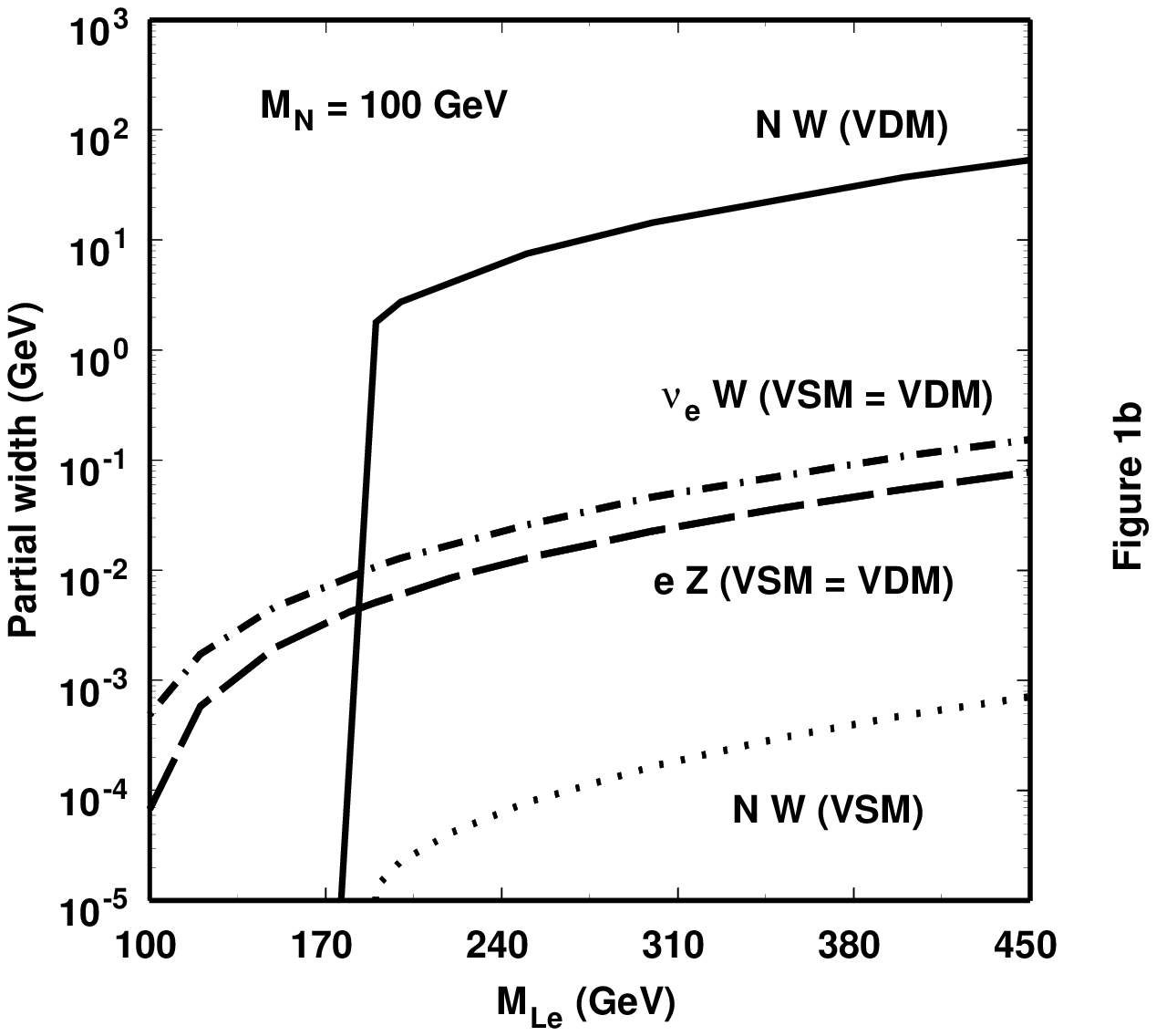}}
\caption{Total heavy lepton width (1a) and partial widths (1b) for VSM and VDM as a function of $M_{L_e}$. The heavy neutrino mass is taken as $M_{N}=100$ GeV.}
\end{figure}

In Figs. 1a and 1b we give the total and partial widths for heavy charged lepton decay as a function of its mass. The enhancement for the VDM model, shown in Fig. 1a, comes from the heavy-to-heavy transition in the channel $L_e \longrightarrow N W $ as shown in Fig 1b and it starts when $ M_{L_e}=M_N +M_W $. This channel is suppressed in the VSM due to the $\sin^2 \theta_{mix}$ factor that is present in the $W-L-N$ vertex, as shown in the last line of table 1. For the VDM, the corresponding term is maximum for this vertex, as shown in table 2,
strongly suppressing the decays $L_e \longrightarrow e + Z$ and $L_e \longrightarrow \nu_e + W$. Of course, if the neutral heavy lepton mass increases, this effect is smaller. We can have important model dependence for the signatures of charged heavy lepton decays. \par

\begin{figure}
\resizebox{1.2\hsize}{!}{\includegraphics*{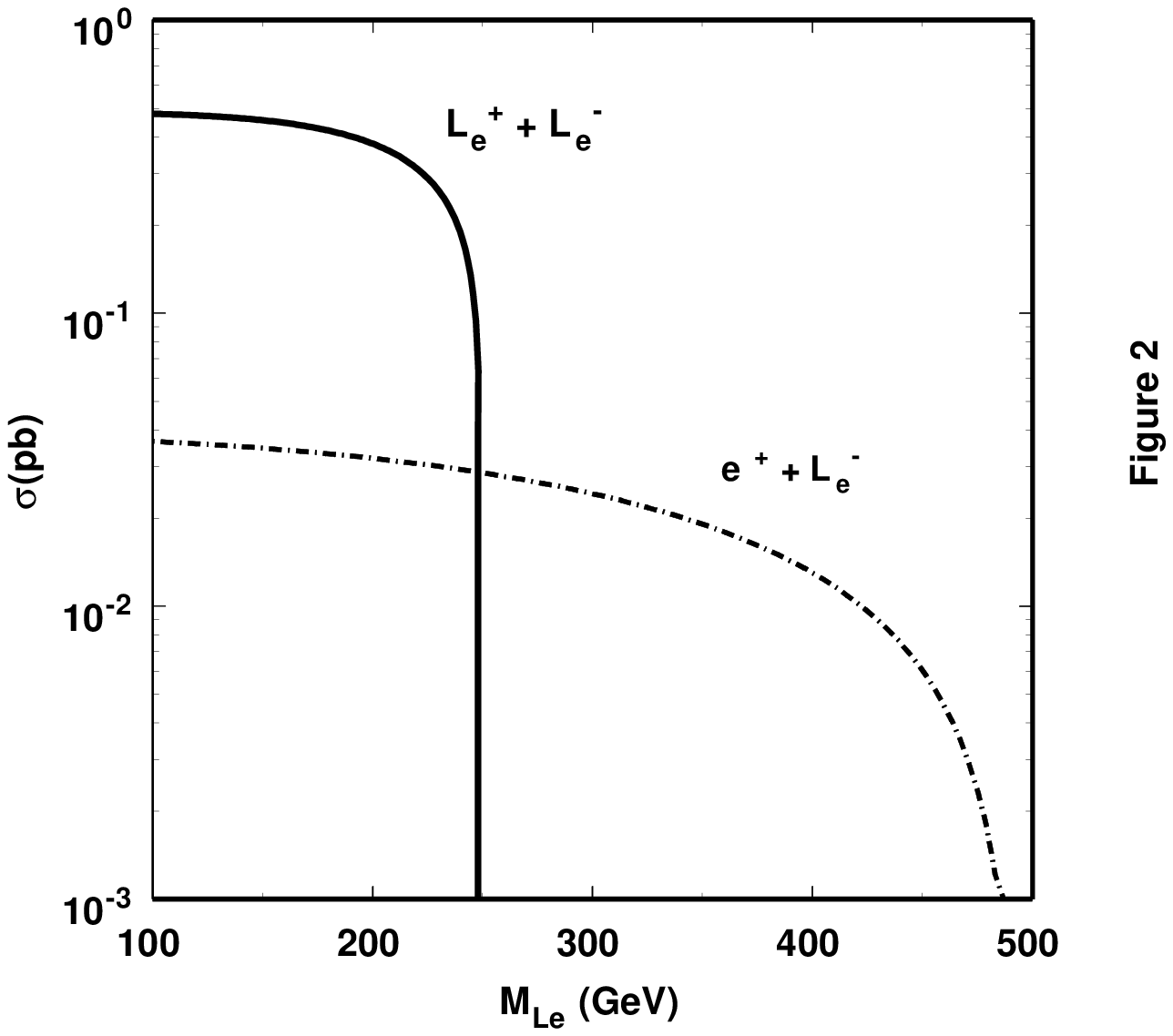}}
\caption{Total cross section for single and pair production of heavy leptons in $e^- + e^+$ at $\sqrt s= 500$ GeV ($\sin^2 \theta_{mix}=0.0052$) for VSM and VDM.}
\end{figure}

For pair production of heavy charged leptons we have three diagrams, but the result is dominated by photon exchange. This is the leading process for masses up to ${\sqrt s}/2$. The single heavy lepton production gives a significative cross section for masses $ > \sqrt{s}/2$ and depends strongly on the mixing parameters. The corresponding total cross sections for real pair and single charged heavy lepton production are given in Fig. 2. Both the VDM and the VSM give the same results. In order to improve our calculations as close as possible to experimental situations, we have employed the following detector cuts for final state particles: a minimum $5$ GeV energy cut for all final leptons; an angular cut $\vert cos \, \theta \vert \geq 0.995$ for all particles, except for the final gauge bosons, relative to the initial electron beam direction and an invariant electron-positron mass cut $\geq 5$ GeV.\par

\begin{figure}
\resizebox{0.75\hsize}{!}{\includegraphics*{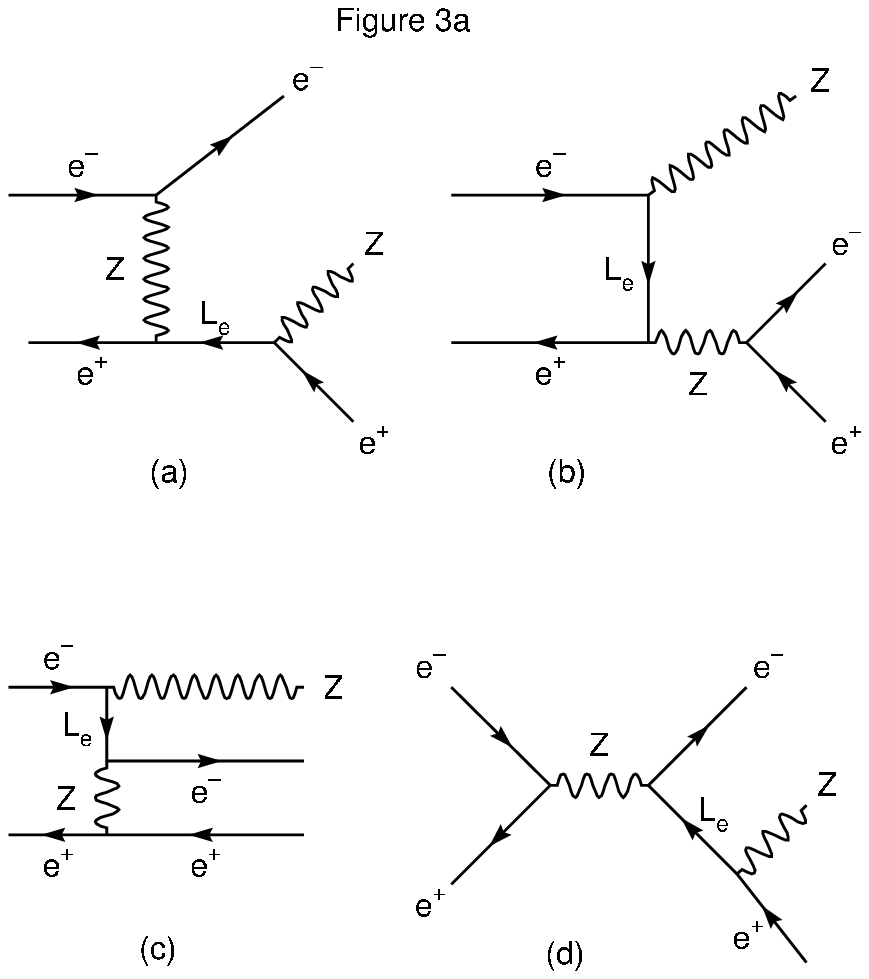}}
\resizebox{0.75\hsize}{!}{\includegraphics*{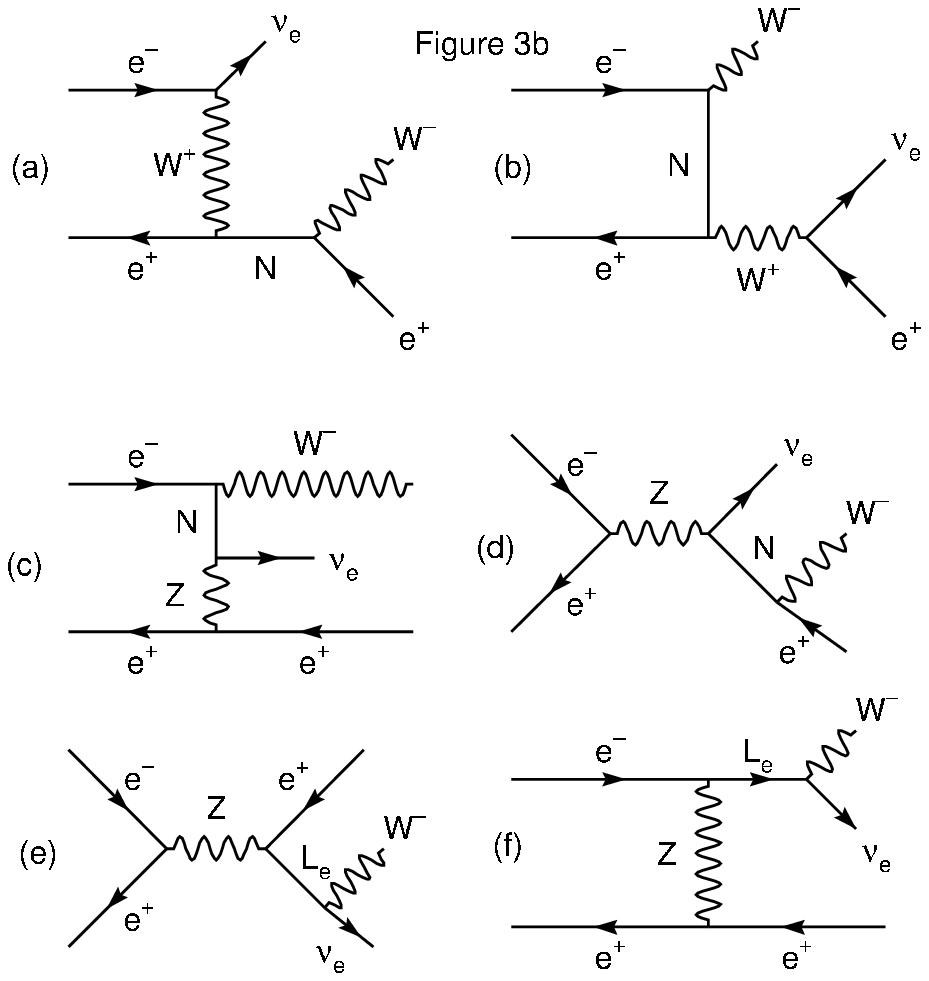}}
\caption{General Feynman graphs for charged heavy lepton contribution in $e^- + e^+ \longrightarrow e^{\pm} + e^{\mp} + Z$ (3a) (8 dominant diagrams) and $e^- + e^+  \longrightarrow e^{+}+ \nu_e \, + W^- $ (3b) (6 dominant diagrams) with $M_N=100$ GeV.}
\end{figure}

For single heavy lepton production one can easily compute cross sections for the three particle final states $e e Z$ and $\nu e W$ by multiplying the results of Fig. 2 by the corresponding branching fractions. As we are mainly interested in comparing different signatures with the SM background, we have performed the full first order calculation. The Feynman diagrams for the signal are given in Figs. 3a and 3b. We notice that in Fig. 3a, each diagram appears twice. For the SM background we have sixteen diagrams for the first channel and twelve diagrams for the second. This procedure of including all diagrams, allows us to have a reliable result not only for cross sections but also for final state particle distributions \cite{PIL}. These large number of diagrams can be taken into account by employing high energy algebraic programs, such as CalcHep \cite{HEP}. The resulting cross sections as a function of $M_L$ for the processes  $e^- + e^+ \longrightarrow e^{\pm} + e^{\mp}+ Z $ and
$e^+ + e^- \longrightarrow e^+(e^-) + \nu_e (\bar \nu_e) +W^-(W^+) $ are shown in Figs 4a and  4b. The curves labeled VDM and VSM correspond to the signal terms only. The first channel has only the charged heavy lepton signal (with both charges) but the second channel has both neutral and charged heavy leptons (with only one charge) contributions, as shown by Figs. 3a and 3b. We show in Fig. 4b the role of the heavy neutrino contribution to the signal total cross section, by computing separately the heavy charged lepton contribution only. Since the dominant process is the real production of heavy leptons, our results are gauge invariant. These are the main signatures for a single heavy charged lepton with mass in the $250-500$ GeV range. \par

\begin{figure}
\resizebox{1.0\hsize}{!}{\includegraphics*{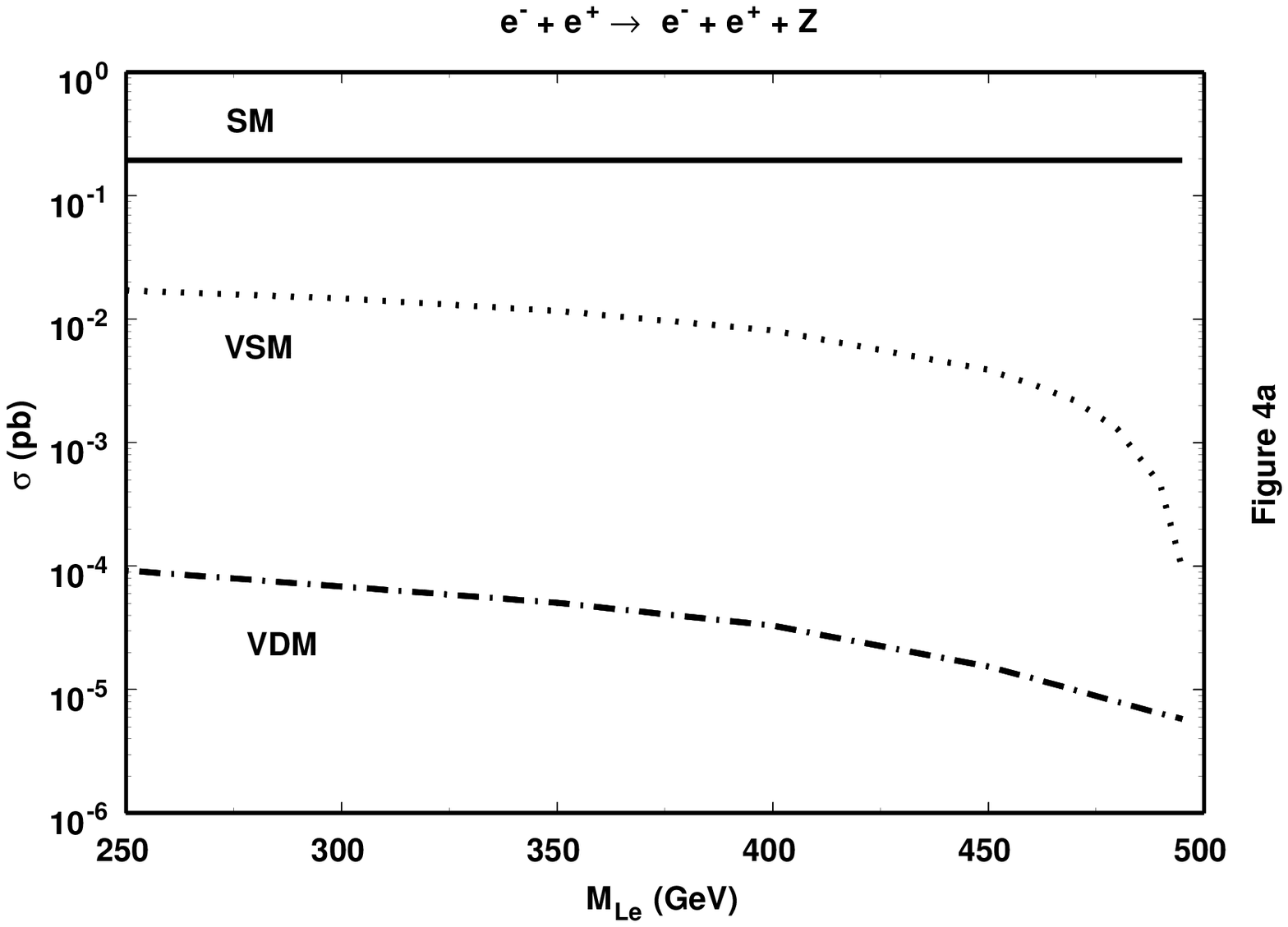}}
\resizebox{1.0\hsize}{!}{\includegraphics*{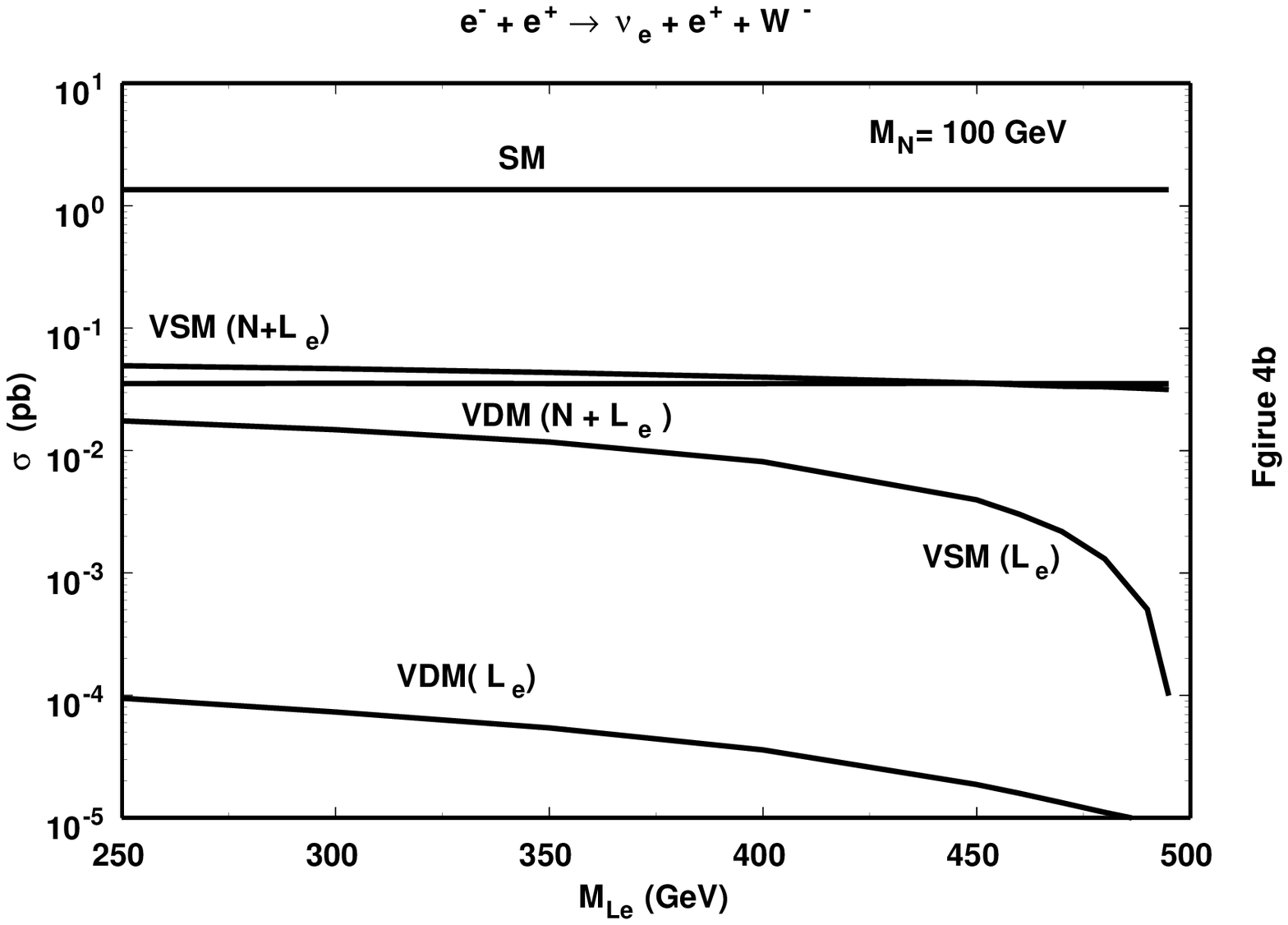}}
\caption{Total cross sections for $e^- + e^+ \longrightarrow e^{\pm} + e^{\mp}+ Z $ (4a) and
$e^- + e^+ \longrightarrow e^+(e^-) + \nu_e (\bar \nu_e) +W^-(W^+) $ (4b) at $\sqrt s = 500$ GeV with $M_N=100$ GeV. In fig. 4b we have separated the pure signal $ N+ L_e $ from the pure signal $L_e$ contributions.}
\end{figure}

For pair production with masses in the $100-250$ GeV range we must look for final states with four particles. For example in Fig. 5 we show the cross sections as a function of $M_L$ for the channels 
$e^+(e^-) + \nu_e (\bar \nu_e) + W^-(W^+) +Z $ and $e^{\pm} + e^{\mp}+ Z +Z$. The curves labeled VSM and VDM, include only the signal from heavy neutral and charged leptons. We notice that the signal is above the SM. For the VDM model and $M_N$= 100 GeV, the contribution of new heavy leptons to the total cross section is the same as in the VSM up to the region $ M_{L_e}=M_N +M_W $. Then the dominant decay mode is $ L_e \longrightarrow N + W $ and the other channels become suppressed.\par

\begin{figure}
\resizebox{1.0\hsize}{!}{\includegraphics*{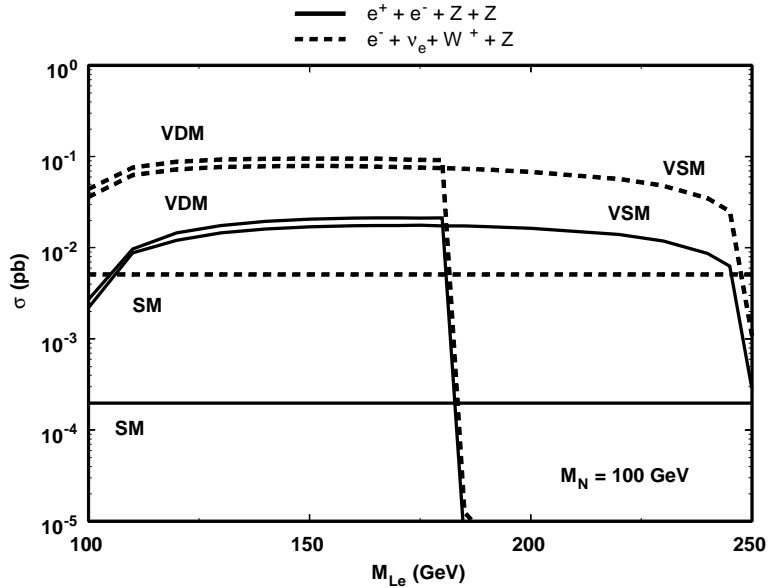}}
\caption{Total cross sections for $e^- + e^+ \longrightarrow e^{\pm} + e^{\mp}+ Z + Z$ and $e^+ + e^- \longrightarrow e^+(e^-) + \nu_e (\bar \nu_e) +W^-(W^+) + Z$ at $\sqrt s = 500$ GeV with $M_N=100$ GeV.}
\end{figure}

 In the experimental search for heavy charged leptons a fundamental point is the choice of distributions that can enhance the signal from the SM background. A simple distribution is the final state invariant mass $M_{eZ}$, as shown in Fig. 6 for the $e^- e^+ Z$ channel, for the complete VSM extension (signal + background). This distribution requires the experimental reconstruction of the Z hadronic jets. There are intrinsic difficulties and uncertainties in the experimental and theoretical  reconstruction of hadronic events. \par

\begin{figure}
\resizebox{1.2\hsize}{!}{\includegraphics*{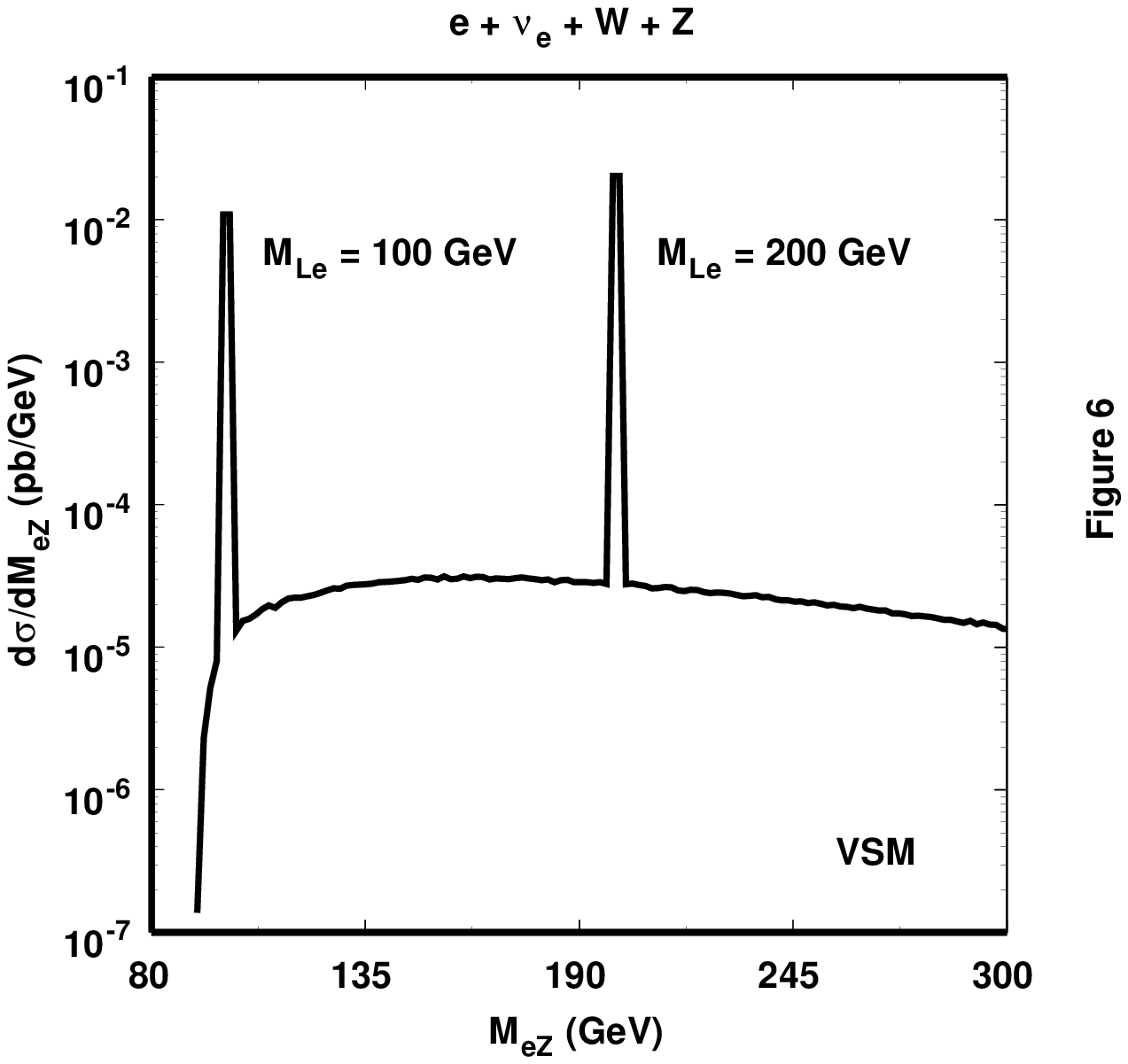}}
\caption{Reconstruction of the invariant mass $M_{eZ}$ for $e^+ +\nu_e + W^- + Z$ at $\sqrt s = 500$ GeV with $M_N=100$ GeV for the complete VSM extension (signal + background).}
\end{figure}

In order to account for  the hadronic effects in our study we have employed the package PYTHIA 6.1 \cite{PIT} to generate the quark hadronization and decay from real W and Z production.  
\par
 An important point that we call attention in this paper is the kinematical relation for a primary electron (positron) produced with a  $L_e$ that must satisfy:
\begin{equation}
E_{e} \pm \Delta E={(s-({M_{L_e} \mp \Gamma_{L_e}})^2)\over{ 2 \sqrt{s}}}
\end{equation}
\par
For a sharp $\Gamma_{L_e}$ we will have a sharp energy distribution of the primary lepton produced together with the new heavy charged lepton.
 This property suggests interesting signatures for heavy charged lepton searches by looking at final state charged lepton distributions. This is far more effective than the reconstruction of invariant masses or hadron jets from $W$ and $Z$ hadronic decays. Analyzing the Fig. 7, for $\nu_e + e^+ + W^-$ channel, we can see that the peak $47.5$ GeV ($240$ GeV) at the final electron (neutrino) energy distribution is related to the $450$ GeV ($100$ GeV) heavy charged (neutral) mass.\par

\begin{figure}
\resizebox{1.2\hsize}{!}{\includegraphics*{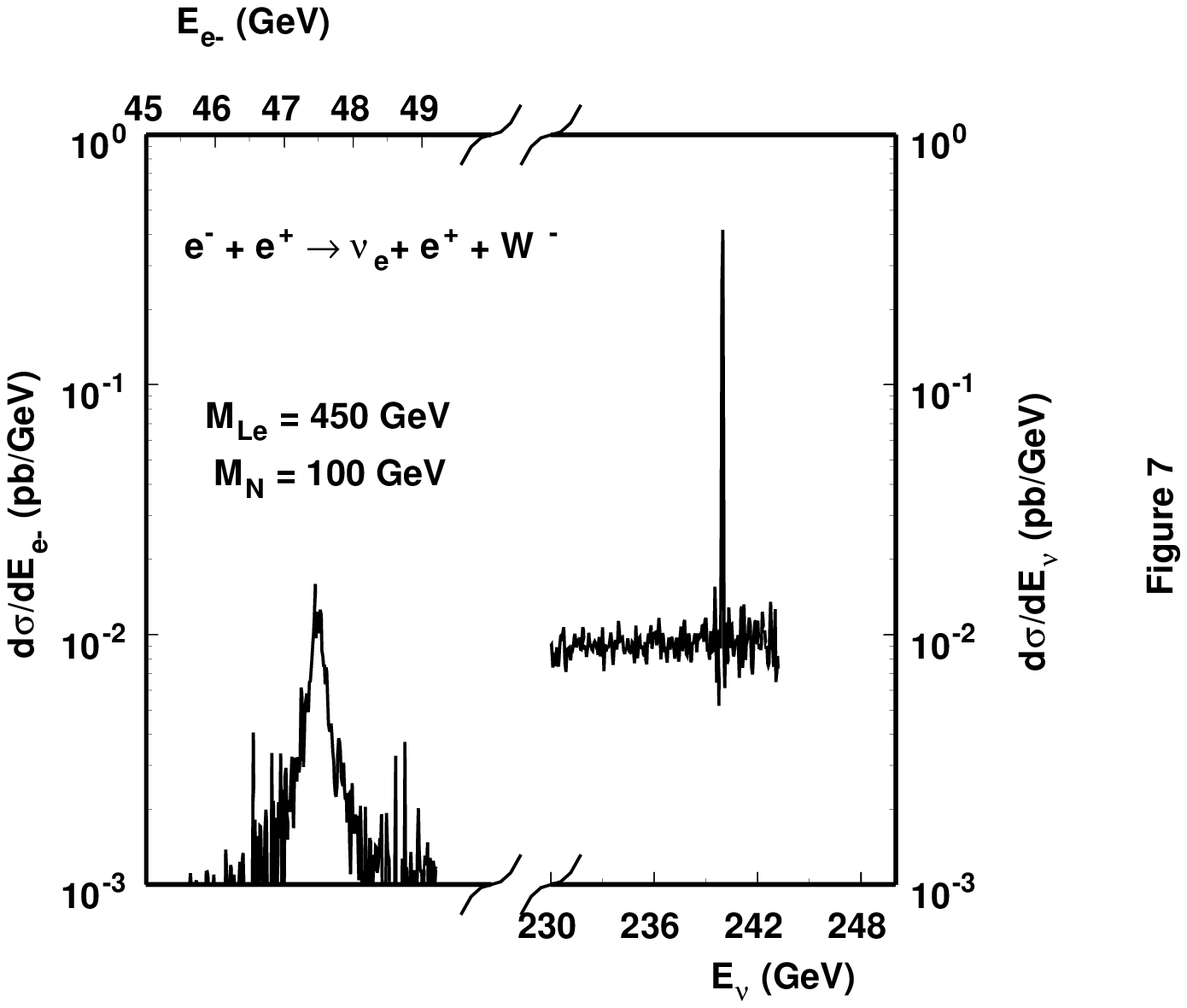}}
\caption{Final lepton energy  and invariant mass distributions in the $ e^+ + \nu_e + W^-$ channel at $\sqrt s = 500$ GeV, for $M_N=100$ GeV and $M_{L_e}=$ 450 GeV.}
\end{figure}

 We also have reconstructed the final state electron energy distribution in the $e^{\pm} + e^{\mp}+$ hadrons  channel. The background and $Z$ hadronization were generated with the PYTHIA package as mentioned above.
A clear separation between signal and SM background is given by 2-dimensional plots instead of usual 1-dimensional plots.
 We have considered the 2-dimensional phase space plot of the electron and positron energy in the $e^{\pm} + e^{\mp}+$ hadrons  channel. According to equation 4, the electron and positron energy of the events coming from the signal are peaked at $160$ GeV for a $300$ GeV charged lepton mass. This correspond to the two crossed lines in Fig. 8a since the process is symmetric. The other concentration at the inclined line is from the standard model $Z\rightarrow e^+ + e^- $ decay. For $M_{L_e}=450$ GeV the result is given in Fig. 8b with the dense lines of the signal shifted to the corners of the kinematical region. For the $ e^+ + \nu_e + $ hadrons channel the analogous plot is given by the final neutrino missing energy versus the final positron energy. For this channel we have the contribution of both the neutral and charged  heavy leptons, as shown in the Feynman diagrams of Fig. 3b. In this case we have one line in the missing energy that comes from a neutral heavy lepton decay ($M_{N}=100$ GeV) and the other line in the electron energy comes from the charged heavy lepton decay ($M_{L_e}=450$ GeV). The result is given in Fig. 9 which is a detail of the allowed phase space. We call attention to the fact that the heavy neutral lepton line is more defined than the heavy charged lepton line. This is due to the higher cross section for the neutral heavy lepton, as shown in Fig. 4b and to the different widths. For $M_{L_e}=300$ GeV we have a similar figure with an horizontal line at the same position since we maintain $M_N=100$ GeV and a vertical line at $E_e=160$ GeV.\par

\begin{figure}
\resizebox{1.2\hsize}{!}{\includegraphics*{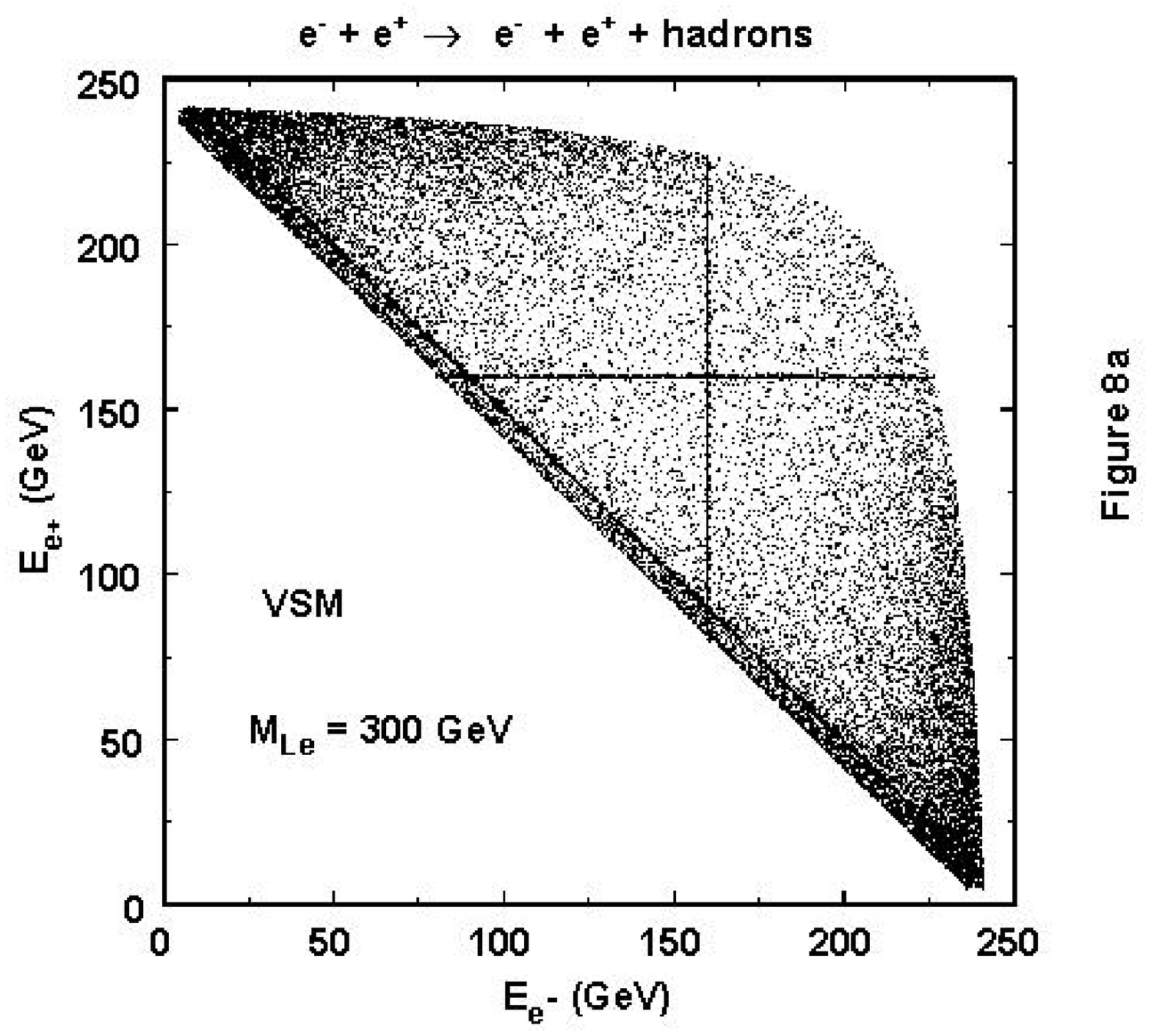}}
\resizebox{1.2\hsize}{!}{\includegraphics*{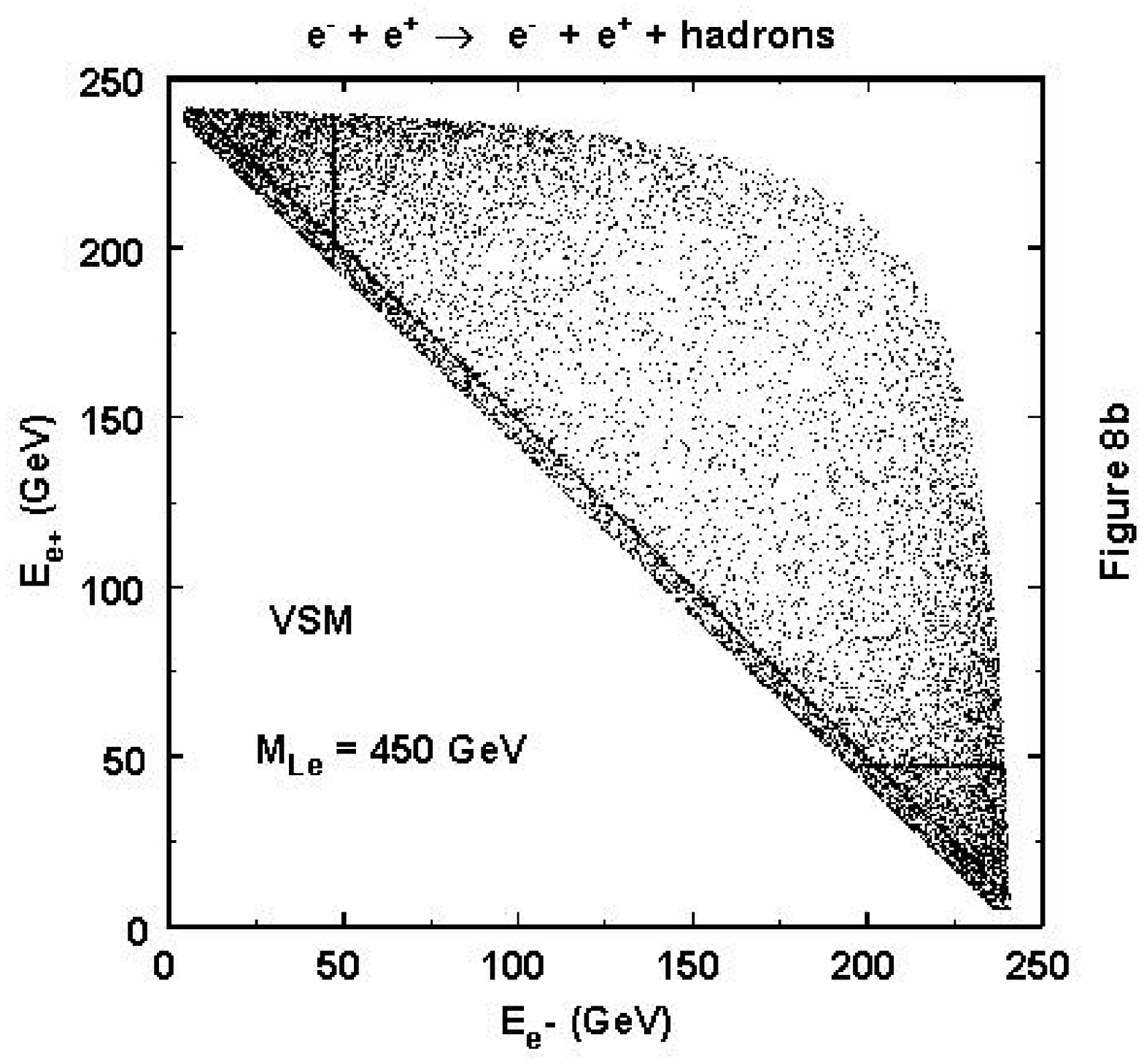}}
\caption{Positron energy versus electron energy event plot in the $e^{\pm} + e^{\mp}+$ hadrons channel for $M_{Le}= 300$ GeV (8a), and the same distribution for $M_{Le}= 450$ GeV (8b) at $\sqrt s = 500$ GeV for VSM.}
\end{figure}

Besides the general cuts mentioned before, the phase space figures suggest that one can perform additional more restrictive cuts in order to enhance the signal over the background. The numbers of events were obtained for an integral luminosity of $ {\cal {L}}=10^5$pb$^{-1}$ and an upper bound for the mixing angle $\sin^2 \theta_{mix} = 0.005$. Table 3 gives the total number of events for the signal $s$ (VSM) and for the SM background $b$, as well as the statistical significance $s /{\sqrt b}$ after all cuts.
As the planned luminosities could be increased \cite{NLC} by a factor of $3$, one could explore 
even lowers values for the mixing angle.\par

\begin{center}
\begin{table*}
\caption{Number of events after hadronization and selection cuts for signal (s)(VSM) + background (b) (SM), background, significance and energy cuts. 
The cut $E_{e^+} + E_{e^-}(E\!\!\!\slash) > 251$ GeV eliminates the inclined line in phase space corresponding to $Z\rightarrow e^+ + e^- (W \rightarrow e + \nu$).}
\begin{tabular}{|c|c|c|c|c|c|} \hline
\multicolumn{6}{|c|}{$e^+ + e^- \longrightarrow e^+ + e^- +$ hadrons} \\ \hline
 Lepton masses &  $s+b$  &  b     & $s/\sqrt b$ & Cuts & Reconstructed mass \\ \hline
 $M_{L_e}= 300$ GeV &  764 & 251 & 30 & $E_{e^+} + E_{e^-} > 251$ GeV & $M_{L_e}$  \\
 &         &      &       & $158 < E_{e^{\pm}} < 162$ GeV & \\\hline
% &        &       &         &   &  \\\hline
 $M_{L_e}= 450$ GeV &  396  & 205    & 13 & $E_{e^+} + E_{e^-} > 251$ GeV & $M_{L_e}$ \\
 &         &        &       & $45 < E_{e^{\pm}} < 49$ GeV & \\\hline\hline\hline
%                &         &        &       &        & \\ \hline\hline\hline
\multicolumn{6}{|c|}{$e^+ + e^- \longrightarrow e^+ + \nu_e +$ hadrons} \\ \hline
 $M_{L_e}= 300$ GeV & 1357 & 599 & 31 & $E_{e^+} + E\!\!\!\slash > 251$ GeV & $M_{L_e}$ \\
 $M_N= 100$ GeV &         &        &        & $158 < E_{e^+} < 162$ GeV & \\\hline 
%                &         &        &       &                                & \\ \hline
 $M_{L_e}= 450$ GeV & 477 & 218 & 18 & $E_{e^+} + E\!\!\!\slash > 251$ GeV & $M_{L_e}$\\
 $M_N= 100$ GeV &         &        &        & $45 < E_{e^+} < 49$ GeV & \\\hline
%                &         &        &       &                                & \\ \hline
 $M_{L_e}= 300$ GeV &  2582   & 599 & 81 & $E_{e^+} + E\!\!\!\slash > 251$ GeV  & \\
 $M_N= 100$ GeV &         &        &        & $239 < E\!\!\!\slash < 241$ GeV & $M_N$\\\hline
%                &         &        &       &                                & \\ \hline
 $M_{L_e}= 450$ GeV & 2515 & 599 & 78 & $E_{e^+} + E\!\!\!\slash > 251$ GeV & \\
 $M_N= 100$ GeV &         &        &        & $239 < E\!\!\!\slash < 241$ GeV & $M_N$ \\\hline
%                &         &        &       &                                & \\ \hline
\end{tabular}
\end{table*}
\end{center}

\begin{figure}
\resizebox{1.2\hsize}{!}{\includegraphics*{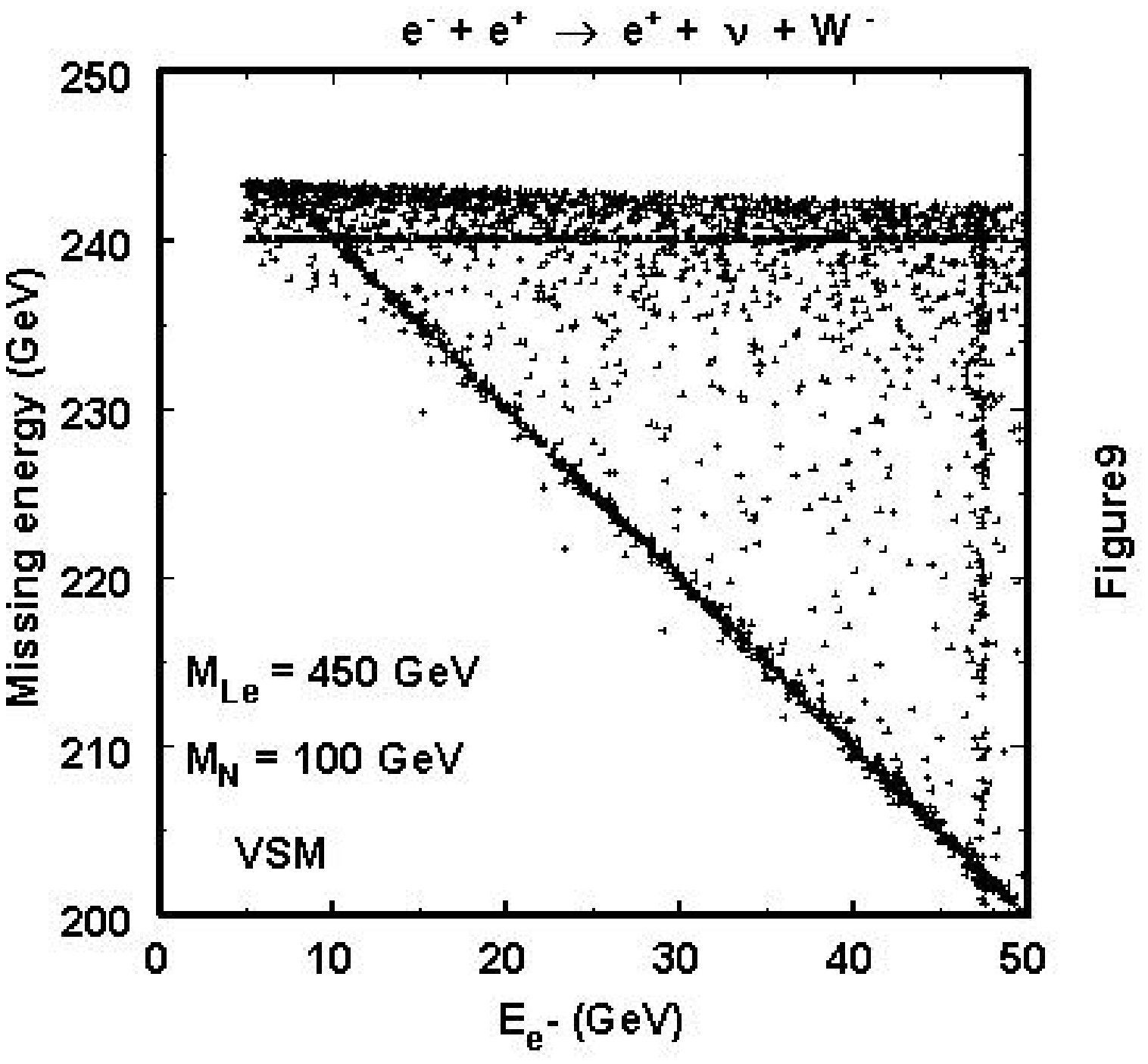}}
\caption{Signal region of interest of the phase space for electron energy versus missing energy event plot in the $e^+ + \nu_e + hadrons$ channel for $M_{L_e}=$ 450 GeV at $\sqrt s = 500$ GeV with $M_N=100$ GeV for VSM.}
\end{figure}

For the $e^{\pm} + e^{\mp}+$ hadrons  channel, the resulting number of events after all cuts is given in Fig. 10a for  an integrated luminosity of $10^5$pb$^{-1}$. The analogous result for the $e^+ + \nu_e  + hadrons $ channel is given in Fig. 10b.\par

\begin{figure}
\resizebox{0.75\hsize}{!}{\includegraphics*{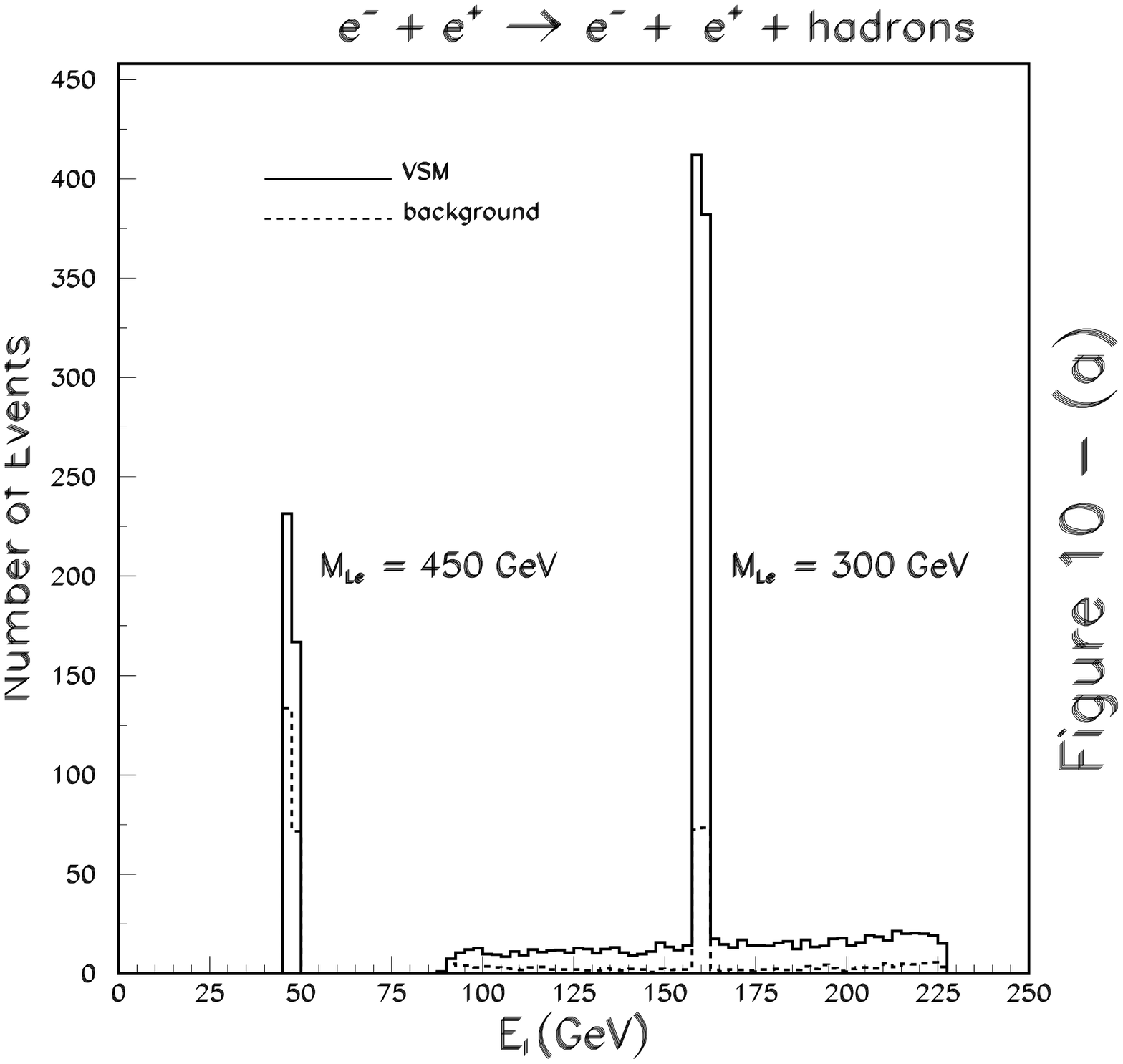}}
\resizebox{0.75\hsize}{!}{\includegraphics*{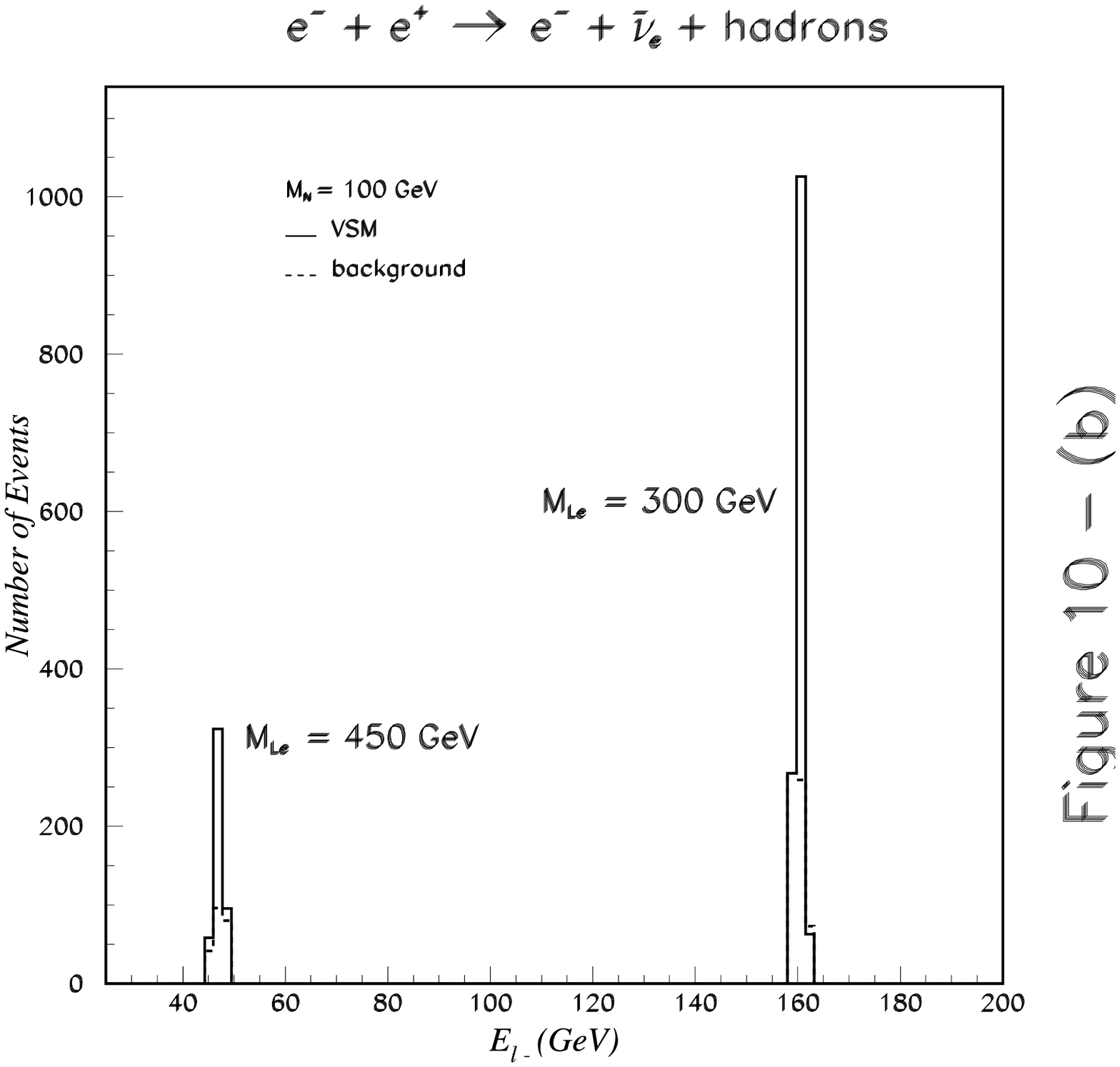}}
\caption{Number of events versus final charged lepton energy in the $e^{\pm} + e^{\mp}+ hadrons$ (10a) and in $ e^+ + \nu_e + hadrons$ (10b) at $\sqrt s = 500$ GeV with $M_N=100$ GeV for VSM.}
\end{figure}

The lepton number violating decay of heavy charged leptons is significant only in VDM. If $M_{L_e} > M_{N} + M_{W}$ and $M_{N} > M_{W}$ we can have processes of the type $e^+ + e^- \longrightarrow e^- + \mu^- +$ hadrons  due to the chain:
\begin{eqnarray}
e^+ + e^-\longrightarrow & & L_e^+ \: e^- \cr
& & \vert  \cr
& & {}^{\displaystyle \longrightarrow N \: W^+}  \cr
& &  \qquad \vert  \cr
& &  \qquad {}^{\displaystyle \longrightarrow \mu^- \: W^+} 
\end{eqnarray}

The total cross section is given in Fig. 11. Here again the single final state electron energy distribution can give information on the mass of the heavy charged lepton as shown in Fig. 12, but with a much larger $ \Gamma_{Le} $ when compared with the VSM. The reconstructed invariant mass $M_{\mu W W}$ gives the same information, but requires the hadronic reconstruction of the invariant masses and/or jets of two Ws with higher uncertainties.

\begin{figure}
\resizebox{1.2\hsize}{!}{\includegraphics*{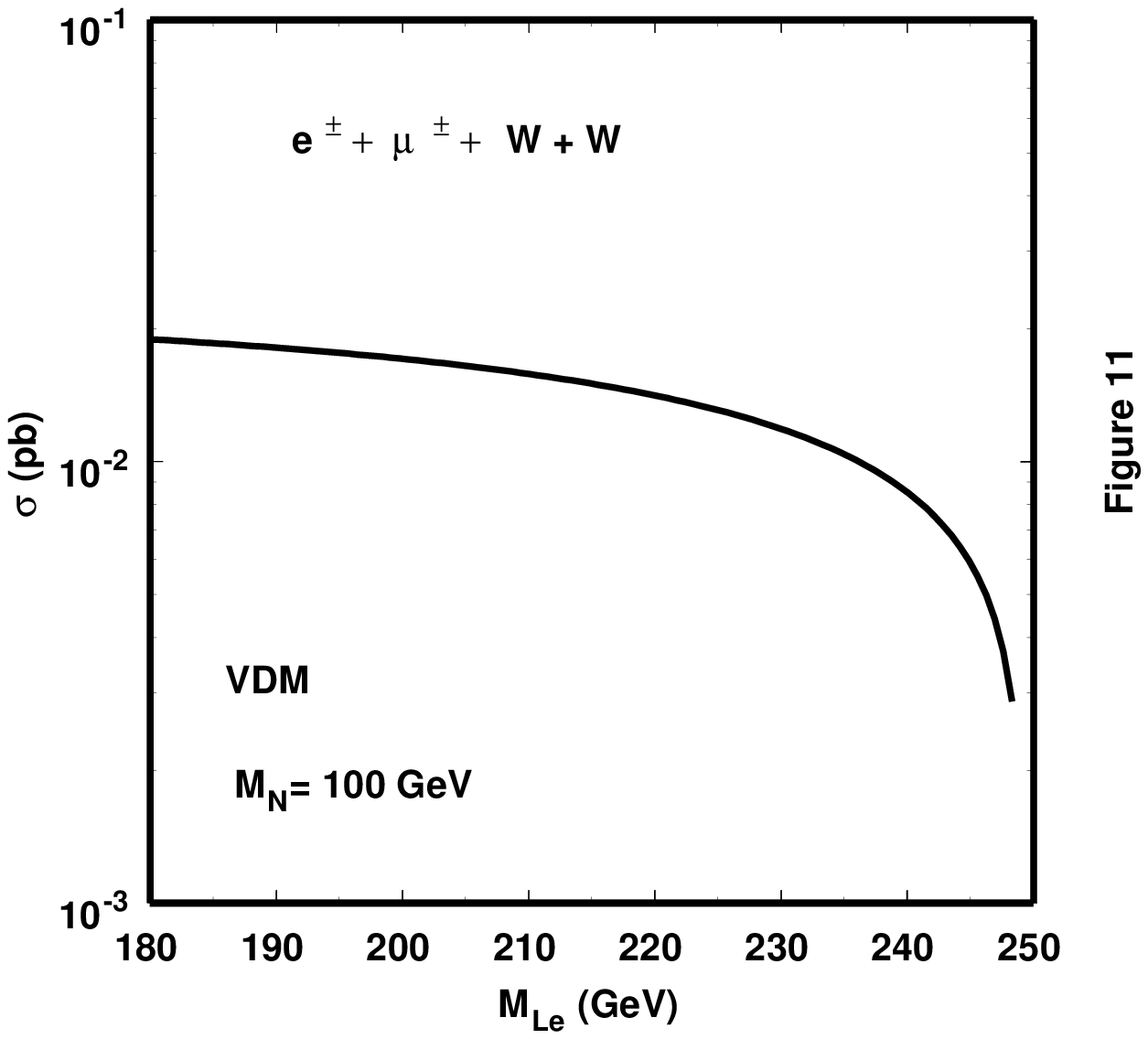}}
\caption{Total cross section for $e^- + e^+ \longrightarrow e^{\pm} + \mu^{\pm} + W + W$ channel at $\sqrt s = 500$ GeV with $M_N=100$ GeV for VDM.}
\end{figure}

\begin{figure}
\resizebox{1.0\hsize}{!}{\includegraphics*{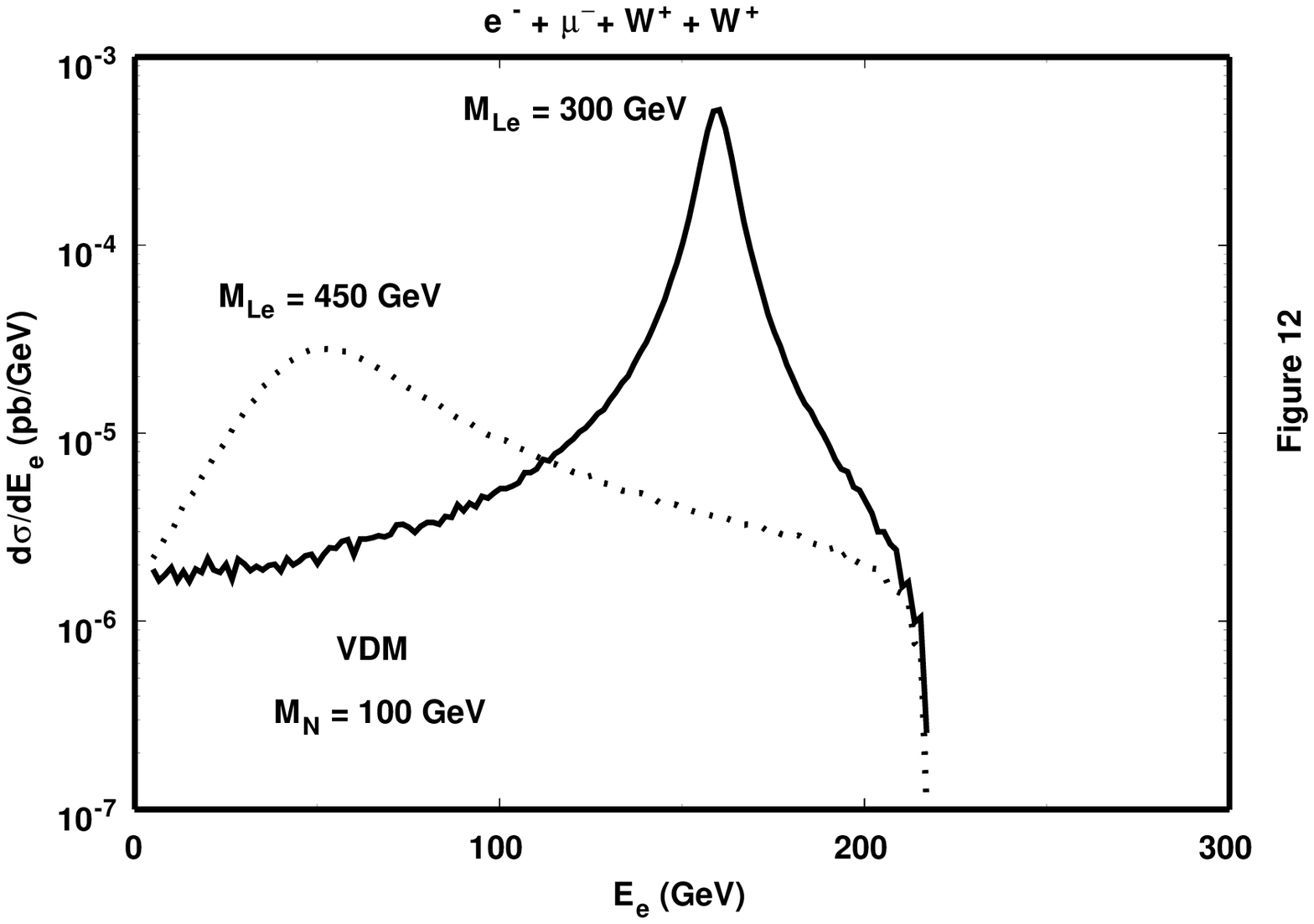}}
\caption{Electron energy distribution in the  $e^{-} + \mu^{-} + W^+ + W^+ $ channel at  $\sqrt s = 500$ GeV with $M_N=100$ GeV for VDM.}
\end{figure}

\section {Conclusions}
In this paper we have investigated different signatures for heavy charged lepton detection at the next electron-positron colliders. We have presented cross sections and distributions for the $ \sqrt {s}= 500$ GeV option for NLC and TESLA colliders, but our conclusions are very similar for higher energies. For vector charged leptons we notice that vector doublets and singlets give the same results if a new heavy neutrino in the same representation has $ M_N \simeq M_{L_e }$. However, if the heavy neutrino mass is smaller than the  new heavy charged lepton, then vector doublets and singlets will show a very different behavior. The decay $L_e \rightarrow N + W$ could
be dominant for vector doublets and suppressed for vector singlets. If the new heavy neutrino is of the Majorana type, we will have the possibility of lepton number violation in the heavy charged lepton decay. This property allows the experimental determination of the Dirac or Majorana nature of heavy neutral leptons.
\par
For new heavy charged leptons with mass up to $M_{L_e} \approx \sqrt {s}/2 $, the dominant process is pair production via photon exchange. Then  the final lepton energy distribution in the processes $e^{\pm} + e^{\mp} + Z$  and  $e^{+}(e^-) + \nu_e(\bar\nu_e) \, + W^-(W^+) $     allows a clear separation from the SM background. If the new heavy charged lepton has a mass in the range $\sqrt {s}/2 < M_{L_e} < \sqrt {s}$ then single production is a feasible process. In this case we call attention to the kinematical properties of the associated primary lepton produced with the new charged lepton, according to the equation 4. Its energy distribution can allow a very precise determination of the new heavy lepton mass without the looses and errors that came from the reconstruction of hadronic events from W and Z decays. This kind of analysis,  in the future $e^- + e^-$ colliders, can lead to much stronger bounds on new heavy lepton parameters if no signal is found .\par

{\it Acknowledgments:} This work was partially supported by the
following Brazilian agencies: CNPq, FUJB and  FAPERJ.

%\bigskip


\begin{thebibliography}{ABC}
\bibitem{GOD} R.M. Godbole, hep-ph/0210196, Presented at 31st International Conference on High Energy Physics (ICHEP 2002), Amsterdam, The Netherlands, 24-31 Jul 2002. 

\bibitem{HEW} J.L. Hewett and T. Rizzo, Phys. Rep. {\bf 183}, 193 (1989); V. Barger, N.G. Deshpande, R.J.N. Phillips and K. Whisnant, Phys. Rev. D {\bf33}, 1912 (1986); 
R.W. Robinett, Phys. Rev. D {\bf33}, 1908 (1986); R. Kitano, Katsuji Yamamoto, Phys. Rev. D {\bf62}, 073007 (2000). 

\bibitem{GON} M.C. Gonzalez-Garcia, Oscar J.P. Eboli, F. Halzen, S.F. Novaes, Phys.Lett. B {\bf 280}, 313 (1992). 

\bibitem{MUR} J. Maalampi and M. Roos Phys. Rep. {\bf186}, 53 (1990); F. Csikor and I. Montvay Phys.Lett. B {\bf 324}, 412 (1994);
Y.A. Coutinho, J.A. Martins Sim\~oes and C.M. Porto, Eur. Phys. J. C {\bf 18}, 779 (2001). 

\bibitem {VIC} F. Pisano and V. Pleitez, Phys. Rev. D {\bf46}, 410 (1992); 
J.E. Cieza Montalvo and M.D. Tonasse Phys. Rev. D {\bf67}, 075022 (2003); J. E. Cieza Montalvo, P. P. de Queiroz Filho, Phys. Rev. D {\bf66}, 055003 (2002).

\bibitem{PDG} D.E. Groom et al., Eur. Phys. Jour. C {\bf 15}, 1 (2000).

\bibitem{ZUB} K. Zuber, Phys. Rep. {\bf 305}, 295 (1998).

\bibitem{HAD} F.M.L. Almeida Jr., Y.A. Coutinho, J.A. Martins Sim\~oes, M.A.B. do Vale, Phys. Rev. D {\bf62}, 075004 (2000);
A. Ali, A.V. Borisov, N.B. Zamorin, hep-ph0112043, Proceedings of 10th Lomonosov Conference on Elementary Particle Physics, Moscow, Russia, 23-29 Aug 2001 and  Eur. Phys. J. C {\bf21}, 123 (2001); O. Panella, M. Cannoni, C. Carimalo, Y.N. Srivastava, Phys. Rev. D {\bf 65}, 035005 (2002). 

\bibitem{LHA} F.M.L. Almeida, J.A. Martins Simoes, A.J. Ramalho, Nucl. Phys. B {\bf347}, 537 (1990); F.M.L. Almeida Jr., Y.A. Coutinho, J.A. Martins Sim\~oes, M.A.B. do Vale, Phys. Rev. D {\bf65}, 115010 (2002);
W. Buchm\"uller and C. Greub, Nuc. Phys. B {\bf 363}, 345 (1991); W. Rodejohann, K. Zuber, Phys. Rev. D {\bf62}, 094017 (2000); Y.A. Coutinho, A.J. Ramalho, Stenio Wulck, Phys. Rev. D {\bf54}, 1215 (1996). 

\bibitem{LEL} F.M.L. Almeida Jr., J.H. Lopes, J.A. Martins Sim\~oes and C.M. Porto, Phys. Rev. 
D {\bf44}, 2836 (1991); F.M.L. Almeida Jr., J.H. Lopes, J.A. Martins Sim\~oes, P. P. Queiroz Filho and A. J. Ramalho, Phys. Rev. D {\bf51}, 5990 (1995);
A. Djouadi Z. Phys. C {\bf63}, 317 (1994); G. Azuelos, A. Djouadi Z. Phys. C {\bf63}, 327 (1994); G. Cveti\u c and C.S. Kim, Phys. Lett. B {\bf461}, 248 (1999), ibid B {\bf471}, 471E (2000).

\bibitem{PHO} J. Peressutti, O.A. Sampayo, Jorge Isidro Aranda, Phys. Rev. D {\bf64}, 073007 (2001); N. Romanenko, Phys. Rev. D, {\bf67}, 033002 (2003). 

\bibitem{NLC} American Linear Collider Group, hep-ex/0106055 v1.


\bibitem{TES} Physics at an $e^+ e^-$ Linear Collider, TESLA Technical Design Report, Part III, Eds. R.-D. Heuer, D. Miller, F. Richard, P.M. Zerwas, March 2001.

\bibitem{MIX} F.M.L. Almeida Jr., Y.A. Coutinho, J.A. Martins Sim\~oes, M.A.B. do Vale, Phys. Rev. D {\bf62}, 075004 (2000); P. Langacker, in  Precision test of the Standard Model, ed. P. Langacker, World Scientific, Singapore, 1995, pg. 883; P. Bamert et al., Phys. Rev. D {\bf54}, 4275 (1996); E. Nardi, E. Roulet and D. Tommasini, Phys. Lett. B {\bf344}, 225 (1995); D. London, in Precision test of the Standard Model, ed. P. Langacker, World Scientific, Singapore, 1995, pg. 951.

\bibitem{VIO} G. Cvetic, C. Dib, C.S. Kim and J.D. Kim, Phys. Rev. D {\bf66}, 034008 (2002). 

\bibitem{PIL} A. Pilaftsis, Phys. Rev. D {\bf65}, 115013 (2002).

\bibitem{HEP} A. Pukhov, E. Boss, M. Dubinin, V. Edneral, V. Ilyin, D.
Kovalenko, A. Krykov, V. Savrin, S. Shichanin and A. Semenov, 
"CompHEP"- a package for evaluation of Feynman diagrams and integration over 
multi-particle phase space. Preprint INP MSU 98-41/542, hep-ph/9908288.

\bibitem{PIT} T. Sj\"ostrand, Comp. Phys. Commun. {\bf82}, 74 (1994).

\end{thebibliography}
\end{document}